\documentclass[12pt,preprint]{aastex}
\usepackage{amssymb}

\begin{document}

\title{High-Resolution Mid-Infrared Morphology of Cygnus A}

\author{James T. Radomski, Robert K. Pi\~{n}a, Christopher Packham, Charles M.
Telesco}
\affil{Department of Astronomy, University of Florida, Gainesville, FL
32611, USA}

\and 

\author{Clive N. Tadhunter}
\affil{Department of Physics and Astronomy, University of Sheffield,
Sheffield, S3 7RH, UK.}

\begin{abstract}
We present subarcsecond resolution mid-infrared images at 10.8 and 18.2 $%
\micron$ of Cygnus A. These images were obtained with the University of
Florida mid-IR camera/spectrometer OSCIR at the Keck II 10-m telescope. Our
data show extended mid-IR emission primarily to the east of the nucleus with
a possible western extension detected after image deconvolution. This
extended emission is closely aligned with the bi-conical structure observed
at optical and near-IR wavelengths by the HST. This emission is consistent
with dust heated from the central engine of Cygnus A. We also marginally
detect large-scale low level emission extending $>$ 1.5 kpc from the nucleus
which may be caused by in-situ star formation, line emission, and/or PAH
contamination within the bandpass of our wide N-band filter.
\end{abstract}


\section{Introduction}

Cygnus A is the most powerful extragalactic radio source in the local
universe (located at a distance of 224 Mpc, H$_{o}=75$ km s$^{-1}$ Mpc$^{-1}$%
) and was among the first extra-galactic radio sources to be optically
identified (Baade \& Minkowski 1954). However, the true nature of the
central engine has remained controversial since that time. Cygnus A is the
prototypical FR II radio galaxy containing a powerful radio jet extending
approximately 80 kpc on either side of the nucleus. In addition, the nuclear
region of Cygnus A may contain a heavily extinguished quasar, hidden from
direct view by up to A$_{V}=$170$\pm $30 magnitudes of extinction as
determined by x-ray observations of Ueno et al. (1994).

Key evidence for the embedded quasar model of Cygnus A was provided through
the detection of broad emission lines by Antonucci, Hurt, \& Kinney (1994)
and Ogle et al. (1997). Additionally, x-ray spectra taken by Ueno et al.
(1994) suggest the presence of an obscured power-law source, entirely
consistent with that expected of a quasar.\ Further evidence for a central
AGN\ is provided by the detection of a bi-conical ionization structure in
the inner 3$\arcsec$, observed at optical and near-IR wavelengths by
Jackson, Tadhunter, \& Sparks (1998) and Tadhunter et al. (1999) respectively
using the HST. The cone axis is aligned to within 15$\degr$ of the radio
axis and, in the near-IR, resembles an edge-brightened bipolar structure
typically observed around young stellar objects (see Velusamy \& Langer
1998).

The detection of an ionization cone and broad emission lines near the
nucleus of Cygnus A provides general support for the so-called `unification
theories of AGN' (i.e. Antonucci \& Miller, 1985). These models account for
the dichotomy in active galaxies in which broad and narrow emission lines
are observed in Type 1 AGN whilst Type 2 AGN display only narrow emission
lines. Unified theories presume that an optically and geometrically thick
dusty torus surrounds the central engine, obscuring the broad emission line
region (BLR) from some lines of sight. The narrow emission line region (NLR)
typically extends over a large volume and hence is less affected by the line
of sight to the central engine. When the torus is viewed face on, the BLR
and NLR from the central engine can be observed. When the torus is viewed
edge on, the BLR is obscured and only narrow emission lines are detected. By
observing Type 2 sources at emission line wavelengths (such as [OIII], H$%
\alpha $ and [NII]), ionization structures have been detected in several
sources (i.e. Schmitt \& Kinney, 1996; Falcke et al., 1998). In many cases
the structures are conical or bi-conical, consistent with ionization by a
central source collimated by a surrounding torus. The detection of an
ionization cone at optical wavelengths along with detection of broad
emission lines lend support to the idea that the active core of Cygnus A is
obscured from direct view by a torus.

In this paper we present subarcsecond resolution 10 and 18 $\micron$ images
of the central $\sim $ 8 kpc of Cygnus A. These images show evidence of the
bi-conical structures seen in near-IR\ observations with HST (Tadhunter et
al. 1999). We discuss these observations and results in Sections 2 and 3.
The origin of the mid-infrared emission is evaluated on nuclear (sub-kpc)
and large (1.5-2 kpc) scales in Section 4, while Section 5 summarizes our
conclusions.

\section{Observations}

Observations of Cygnus A were made on May 9, 1998 using the University of
Florida mid-infrared camera/spectrometer OSCIR on the Keck II 10-m
telescope. OSCIR uses a 128 x 128 Si:As Blocked Impurity Band detector
developed by Boeing. On the Keck II 10-m telescope OSCIR has a plate scale
of 0\farcs062 pixel$^{-1}$ with the f/40 chopping secondary, corresponding
to a field of view of \ 7\farcs9 x 7\farcs9. Images were obtained in the N ($%
\lambda _{o}$=10.8 $\micron$,\ $\Delta \lambda $=5.2 $\micron$) and IHW18 ($%
\lambda _{o}$=18.2 $\micron$,\ $\Delta \lambda $=1.7 $\micron$) filters
using a standard chop/nod technique to remove sky background and thermal
emission from the telescope. The chopper throw was 8$\arcsec$ in declination
at a frequency of 4 Hz and the telescope was beam switched every 30 seconds.
All observations were guided using the off-axis guider. OSCIR was mounted at one of the Keck-II ``bent cassegrain" ports. Due to the
alt-az mount of the Keck-II telescope, any instrument directly attached to it
will see a fixed pupil and a field orientation which rotates as the telescope
tracks. In order to fix the field as seen by the detector array, OSCIR was
mounted on an instrument counter-rotator. However, this does produce a rotation
of the pupil as the telescope tracks, and correspondingly a rotation of the Keck
hexagonal diffraction pattern on the detector array.

Cygnus A was observed for a total on-source integration time of 240 seconds
(480 chopped) at N and 180 seconds (360 chopped) at IHW18. Observations of $%
\gamma $\ Aql were taken directly before Cygnus A for flux calibration.
Absolute calibration of\ $\gamma $\ Aql was achieved using a stellar model
based on spectral irradiance models of similar K3II stars by Cohen (1995)
adjusted for filter and atmospheric transmission. Measurements of other
calibration stars throughout the night showed flux calibration variations of
less than 5\% at N and less than 15\% at IHW18. Observations of $\nu $ Cyg
and $\gamma $\ Aql were used to measure the telescope's point spread
function (PSF). The measured FWHM of $\nu $ Cyg at N was 0\farcs30 based on
a 15 second exposure (30 seconds chopped). Images of $\nu $ Cyg were not
obtained at IHW18 so the flux calibration star $\gamma $\ Aql was used as a
measure of the IHW18 PSF. This yielded a FWHM of 0\farcs4 based on a 30
second exposure (60 seconds chopped). Both measurements lie close to the
theoretical diffraction limit (0\farcs27 at N and 0\farcs38 at IHW18). Short
integrations on $\nu $ Cyg and $\gamma $\ Aql were sufficient for comparison
to Cygnus A due to the stability of the OSCIR/KECK II PSF on May 9, 1998.
Observations of the standard star $\beta $\ Leo taken in 15 second
increments showed that the FWHM varied by $<$ 7\% over a total integration
time of 180 seconds (360 chopped).

Cygnus A images were rotated to place North up and East left in
post-processing. Images of the PSF stars $\nu $ Cyg and $\gamma $\ Aql were
also rotated -14.1$\degr$ and 37.8$\degr$ respectively to match the position
angle of the Keck hexagonal diffraction pattern as projected on the detector
array when Cygnus A was observed. In addition, PSF images
were rotationally ``smeared'' to account for the slight rotation of the
pupil ($\sim$ 3$\degr$) during the exposure times of Cygnus A. These
images were then used to deconvolve the Keck II telescope's PSF from
the Cygnus A images at 10 $\micron$.

\section{Results}

Figure 1 shows our N band image of the central 2 kpc ($\sim 2\arcsec$) of
Cygnus A. Figure 2 shows the same analysis of Cygnus A at IHW18.\ Both
figures clearly show extended emission detected primarily to the east of the
nucleus. This extended emission can be separated into two distinct regions
located northeast and southeast of the central source.\ These regions are
closely aligned with the bi-conical structure (opening angle $\sim $116$\degr$\ ) observed at optical wavelengths by Jackson, Tadhunter, \& Sparks (1998)
and at near-IR wavelengths by Tadhunter et al. (1999). However, at this
resolution the raw data shows little evidence of the western component of
the bi-cone (Figure 3a). Using a deconvolution method from R. K. Pi\~{n}a
2001 (in preparation), the PSF star $\nu $ Cyg (Figure 3b) was deconvolved
from the raw N-band image. Deconvolution using the PSF star\ $\gamma $\ Aql
was also performed with similar results.\ The final deconvolution is shown
in Figure 3c. This figure shows a morphology which is in close agreement
with the structure of the HST 2.0 $\micron$\ near-IR data seen in Figure 3d.
Figure 4\ shows that the deconvolved structure is also very similar to the
inner morphology of the [OI] emission map of Jackson, Tadhunter, \& Sparks
(1998). The [OI] data also implies a more parabolic shape to the ionization
cone structure. In addition, the deconvolved data shows that most of the 10 $%
\micron$ emission arises from the SE-NW limb of the bi-cone. This is similar
to the results of Tadhunter et al. (2000) in which the SE-NW limb also
dominates in near-IR polarization images. Due to low signal-to-noise,
deconvolution was not performed on the IHW18 data.

Table 1 shows our flux measurements of Cygnus A in comparison with other
observation from the literature. These values are also plotted in Figure 5.
Comparing our results with that of IRAS\ shows that while we see $\sim 100\%$
of the predicted 10 $\micron$\ flux, we only detect about 60\% of the
predicted 18$\micron$ flux (based on a simple linear fit). This implies that
most of the warm dust ($<$ 12 $\micron$) lies within the central 2$\arcsec$
region of the galaxy while the cooler dust ($>$ 18 $\micron$) extends
farther out. These results are roughly consistent with the 20 $\micron$ ISO
data of Haas et al. (1998) which measured 816 mJy using a 23$\arcsec$
aperture but does not explain their 12.8 $\micron$ measurement of 485 mJy.
It is unclear why the 12.8 $\micron$\ ISO\ data differs so greatly from the
IRAS data taken in a larger aperture as well as other 10 $\micron$
measurements from the literature.

In addition to the sub-kpc structure of Cygnus A, we also marginally detect
large scale mid-IR\ emission $>1.5$ kpc from the nucleus within the
ionization cone. Flux is detected within both cones at N, while low level
emission at IHW18 is detected only in the southeastern cone. This
large-scale emission can be seen at a 2-3 sigma level in 0\farcs5 gaussian
smoothed 10 and 18 $\micron$ images in Figure 6. Figure 7 shows the
large-scale N-band emission to be approximately coincident with [OIII]
emission located within the ionization cone of Cygnus A. Measurements of
this emission were taken with \ 2$\arcsec$ x 4$\arcsec$ rectangular beam (see Table
1). Subtracting off the 2$\arcsec$ flux measurement of the core leaves us
with $\sim 22\pm 2$ mJy at N and $\sim 90\pm 10$ mJy at IHW18 over a $\sim$ 5 arcsec%
$^{2}$ region.

\subsection{Temperature and Optical Depth}

Temperature and emission optical depth maps from simple radiative transfer
analysis provide a good first-order estimate of the sources of grain heating
as well as the relative density of warm grains (Tresch-Fienberg et al.
1987). Flux maps of the nuclear region of Cygnus A were created by
convolving images at N with the IHW18 PSF and vice-versa to attain the same
resolution. Temperature and emission optical depth estimates were obtained
pixel by pixel by solving the equation of radiative transfer ($F_{\nu
}=\Omega \tau B_{\nu }(T)$) at two wavelengths assuming the optically thin
approximation ($\tau $ $\ll $ 1). Where $F_{\nu }$ is the observed flux
density at frequency $\nu $, $\Omega $ is the solid angle of each pixel, $%
B_{\nu }(T)$ is the Planck function evaluated at frequency $\nu $\ and
temperature T, and $\tau $ is the emission optical depth. The frequency
dependence of dust grain emission efficiency in the mid-IR is approximated
as $Q(v)\propto v^{1}$.

Since no astrometric calibration was performed due to the limited field of
view of OSCIR, the peak flux of the convolved N-band image was aligned to
coincide with peak flux of the convolved IHW18 image. In order to determine
the errors due to alignment, a Monte Carlo simulation was done by shifting
the two convolved images with respect to each other up to $\sim $ 0.1 $%
\arcsec$ in all directions. The structure of the temperature map was highly
dependent on the alignment of the two convolved images. Temperature values
were most stable within the SE cone, varying $\pm $5 K. Temperatures in the
core showed a dispersion of approximately $\pm $10 K, while that in the NE\
cone varied up to $\pm $30 K due to low S/N of any extended emission at 18 $%
\micron$ in this area. The emission optical depth map (Figure 8) was much
less dependent on alignment and consistently showed higher optical depths
along the limbs of the SE ionization cone. This is consistent with the
scenario proposed by Tadhunter et al. (1999) that dust has been destroyed or
swept out of the cones by outflows from the central quasar. Lateral
expansion of these outflows may have also caused density enhancements of
dust along the walls of the bi-cone.

\section{Analysis and Discussion}

There are several possible mechanisms that can account for extended mid-IR
emission: in-situ star formation, material heated by shocks, dust heated by
the central engine, polycyclic aromatic hydrocarbon emission (PAH), and
emission lines. Each is considered below and how they relate to the
mid-infrared emission detected on small (sub-kpc) and large scales (1.5-2
kpc).

\subsection{Origin of Sub Kiloparsec Emission}

\subsubsection{Star Formation}

Extended mid-IR emission has been observed to arise from young star
formation in the central regions of galaxies (Telesco 1988). In Cygnus A
however, estimates of the 2-10 keV hard x-ray luminosity relative to the
40-500 $\micron$ far-IR luminosity is $\gtrsim $ 0.1, typical for a galaxy
predominantly powered by AGN activity (Imanishi et al. 2000). Additionally
the 11.3 $\micron$ PAH\ emission is less than would be expected if the
region contains strong star formation. PAH measurements of Imanishi \& Ueno
(2000) place a lower limit of the ratio of 11.3 $\micron$ luminosity to
far-IR luminosity more than an order of magnitude smaller than that found
for galaxies dominated by star formation (Smith, Aitken, \& Roche 1989). \
Finally, optical spectroscopy of \ Cygnus A by Thornton, Stockton, \& Ridgway
(1999) (using a 1\farcs1 slit oriented along the ionization cone axis) show
the equivalent width of the Ca II triplet $\lambda 8498,\lambda 8542$, $%
\lambda 8662$ to be much less than would be expected if star formation
dominated the nucleus (Terlevich, Diaz, \& Terlevich 1990). Thus it is
unlikely that the majority of the mid-IR emission detected arises from star
formation.

\subsubsection{Heating From Central Engine}

Optical spectroscopy by Thornton, Stockton, \& Ridgway (1999) also detected
[Ar XI] $\lambda 6917$ and [Fe XI] $\lambda 7889$ in the southeastern
component of Cygnus A. High-ionization lines such as these indicate either
photoionization by a continuum source extending to the far-UV or transient
heating by high speed shocks to $\sim $ 2 $\times $ 10$^{6}$ K (e.g.
Osterbrock \& Fulbright 1996). \ Thornton used the line flux ratios of H$%
_{2} $ (F[$\upsilon =1-0$ S(3)]/F[$\upsilon =1-0$ S(1)], Kawara, Nishida \&
Gregory 1990) as well as the ratio of H$_{2}$ and [O I] emitting regions
(Mouri et al. 1989) to rule out shock heating and propose that the most
likely cause for the high-ionization lines was due to x-ray heating from a
central quasar.

The central quasar may also be responsible for the extended mid-IR emission.
Based on color temperature maps of the nuclear region (see section 3.1), we
estimate that dust reaches a T $\sim $ 150$\pm 10$ K up to 500 pc from the
central source. Assuming a uniform dust distribution, a first-order
determination of the size of the region that could be heated by a central
source can be made. The equilibrium temperature of dust in a strong UV field
(Dopita et al. 1998) is given by

\begin{equation}
T\sim 900L_{10}^{0.22}r_{pc}^{-0.44}a^{-0.22}
\end{equation}
where $L_{10}$ is the luminosity of the central source in units of 10$^{10}$
L$_{\sun}$, $r_{pc}$ is the radius from the source in parsecs, and $a$ is
the average grain radius in units of 0.1 $\micron$. \ 

Estimating the true luminosity of an embedded source such as Cygnus A is
difficult due to the high levels of dust obscuration towards the central
source. Haas et al. (1998) estimated a luminosity L$_{1-1000\micron}=$ 4.7 $%
\times $ 10$^{11}$\ L$_{\sun}$ for Cyg A from ISO measurements. Ward et al.
(1991) however, estimated the total luminosity of Cyg A to be as high as $%
\sim $ \ 1.6 $\times $ 10$^{12}$\ L$_{\sun}$. Using Balmer line fluxes in a 5%
$\arcsec$ aperture Stockton, Ridgway, \& Lilly (1994) calculated a bolometric
luminosity of Cyg A of \ L$_{Bol}=$ 3.3 $\times $ 10$^{11}$\ L$_{\sun}$.
However, they also noted that this calculation assumed that the gas in the
central region had a covering factor of unity and hence represents a lower
limit that could easily be an underestimate of the true luminosity by a
factor of 10 or more. Therefore the true luminosity of Cygnus A could be as
high as 3.3 $\times $ 10$^{12}$\ L$_{\sun}$. \ Using equation (1) and
assuming a central luminosity of $\sim $ 1\ $\times $ 10$^{12}$\ L$_{\sun}$,
dust could be heated to a temperature of 150 K from the central engine up to
a distance of $\sim $ 500 pc for grain sizes $\sim $ 0.1 $\micron$.
Classical interstellar dust is generally considered to be a mixture of
silicate and graphite particles with grain sizes in the range (0.003 - 1 $%
\micron$) (Draine \& Lee 1984). Thus the sub-kpc extended mid-IR emission in
Cyg A is consistent with heating of dust from a central engine.

\subsection{Origin of Large Scale Emission}

Although dust heated from the central engine may account for the mid-IR
emission on sub-kpc scales, it may not entirely account for the low level
large-scale emission\ marginally detected at 10 and 18 $\micron$ (see Figure
6). This large scale emission extends 1.5 - 2 kpc from the nucleus and
covers a total area of $\sim $5 arcsec$^{2}$. Based on the average surface
brightness of this emission in the SE cone we estimate a dust T$\sim $150$%
\pm 5$ K (similar to that found in the nuclear regions).\ Referring back to
equation (1), in order for dust grains at $\geq $1.5 kpc to reach these
temperatures from central heating they would need to be $\leq 0.02$ $\micron$%
. \ In the NE cone where only 10 $\micron$ emission is detected, we estimate
a lower temperature limit of approximately 220$\pm 30$ K which would require
dust grains $\leq 0.003$ $\micron$. \ Though these are only approximations
of course, they do imply that central heating may not be able to entirely
explain mid-IR emission on these large scales.

This raises the possibility that small grains such as PAH's could contribute
to the large scale emission. Previous spectrometry by Imanishi \& Ueno
(2000) reveals that the 11.3 $\micron$ PAH feature is weak in Cygnus A. This measurement was made with a 0\farcs5 slit centered on the
nucleus of the galaxy. However, the orientation of the slit with respect to this
large scale extended emission is unclear. If the slit was oriented
perpendicular to this emission it may have only shown that PAH emission is
weak in the core of Cygnus A and may not necessarily rule out PAH\ emission
farther out in the ionization cone.

In-situ heating from ongoing star formation may also contribute to the
large-scale emission. Though mid-IR emission from star formation is probably
weak in the central few kpc of Cygnus A (see section 4.1.1), it may explain
the low level emission marginally detected in the SE cone at N and IHW18. Jackson, Tadhunter, \& Sparks (1998) detected several compact blue regions
within the inner $\sim $2 kpc of Cygnus A using HST filters between
336-622nm. At least 4 of 8 of these regions coincide with the large-scale
extended emission seen at N and IHW18 in the SE cone (see Figure 7). \ Lynds
et al. (1994), Stockton, Ridgway, \& Lilly (1994), and Jackson all suggest
that these regions trace young star formation similar to young star clusters
seen in NGC 1275 (Holtzman et al. 1992).

Another possible explanation for the large-scale extended emission is mid-IR
line emission. The coincidence between the [OIII] and this emission reveal
that ionizing photons of at least 54.9 eV exist in these regions. This
provides sufficient ionization to produce the 12.814 $\micron$ [NeII] fine
structure line (21.6 eV) which would fall within our N-band filter ($\sim $
8-13 $\micron$). Other mid-IR lines such as [ArIII] and [SIV] could also
contaminate our N-band filter and contribute to this emission. Future mid-IR
spectroscopy with the slit placed along the ionization cone is needed to
explore this possibility.

\section{Conclusions}

The mid-IR morphology of the inner kpc of Cygnus A shows structure
consistent with the bi-cone observed at near-IR wavelengths. The SE portion
of the bi-cone is detected in our raw N and IHW18 band data while the NW
cone is detected in our deconvolved 10 $\micron$ image. Our calculations
suggest that these structures are consistent with heating from a central
engine rather than from local heating from star formation. We also
marginally detect low level extended emission at 10 and 18 $\micron$ on
scales up to 1.5-2 kpc from the nucleus. \ This emission may be a result of
heating from the central engine on small grains, in-situ heating from weak
star formation, line emission such as from [Ne II], PAH emission or any
combination of these. An optical depth map of the region shows the highest
optical depth is located along the edges of the ionization cone and supports
the conclusion that dust in the central regions of Cygnus A is swept up by
outflows, possibly causing density enhancements along the walls of the
bicone.

\acknowledgments
We would like to thank the Florida Space Grant Consortium (Grant No. NASA-NGT5-40107) and the National
Science Foundation for funding which assisted in the completion of this work. In addition, we thank R. Scott Fisher and David Ciardi for their invaluable support and
input. Finally, we wish to extend special thanks to those of Hawaiian ancestry on whose sacred mountain we were privileged to
be guests. Without their generous hospitality, none of the observations presented herein would have been
possible.

Data presented herein were obtained at the W.M. Keck Observatory, which is operated as a scientific partnership among the California Institute of Technology, the University of
California and the National Aeronautics and Space Administration. The Observatory was made possible by the generous
financial support of the W.M. Keck Foundation.

\clearpage

\begin{deluxetable}{cccc}
\footnotesize
\tablecaption{Cygnus A Flux Measurements. \label{tbl-1}}
\tablewidth{0pt}
\tablehead{
\colhead{Filter} & \colhead{Aperture}   & \colhead{Flux Density (mJy)}   & \colhead{Reference} } 
\startdata
N-band & 2$\arcsec$ & 104$\pm 3$\tablenotemark{a} & 1  \\
N-band & 2$\arcsec$ $\times$ 4$\arcsec$\tablenotemark{b} & 126$\pm 4$\tablenotemark{a} & 1  \\
N-band & 5.8$\arcsec$ & 180$\pm 30$ & 2 \\
10 $\micron$ & 6$\arcsec$ & 180$\pm 30$ & 3\\
11.7 $\micron$ & 1.92$\arcsec$ & 122 & 4\\
12 $\micron$ (IRAS) & 30$\arcsec$ $\times$ 90$\arcsec$ & 168$\pm 18$ & 5  \\
12 $\micron$ (ISO) & 23$\arcsec$ & 485 & 6\\
\cutinhead{Long Wavelength}
IHW18 (18 $\micron$) & 2$\arcsec$ & 319$\pm 27$\tablenotemark{a} & 1  \\
IHW18 (18 $\micron$) & 2$\arcsec$ $\times$ 4$\arcsec$\tablenotemark{b}& 409$\pm 38$\tablenotemark{a} & 1  \\
25$\micron$ (IRAS) & 30$\arcsec$ $\times$ 90$\arcsec$ & 912$\pm 16$ & 5 \\
20 $\micron$ (ISO) & 23$\arcsec$ & 816 &6 \\
\enddata
\tablenotetext{a}{Error bars include statistical as well as calibration error ($\sim$ 5\% at N-band and $\sim$ 15\% at IHW18}
\tablenotetext{b}{Measured using a rectangular beam with major axis position angle = 60$\degr$ to incorporate large scale emission}
\tablerefs{(1) this paper, also Radomski et al. (2001), (2) Heckman et al. (1983), (3) Rieke \& Low (1972), (4) Whysong \& Antonucci (2001), (5) Knapp et al. (1990), (6) Haas et al. (1998)}
\end{deluxetable}

\clearpage

\begin{figure}[tbp]
\plotone{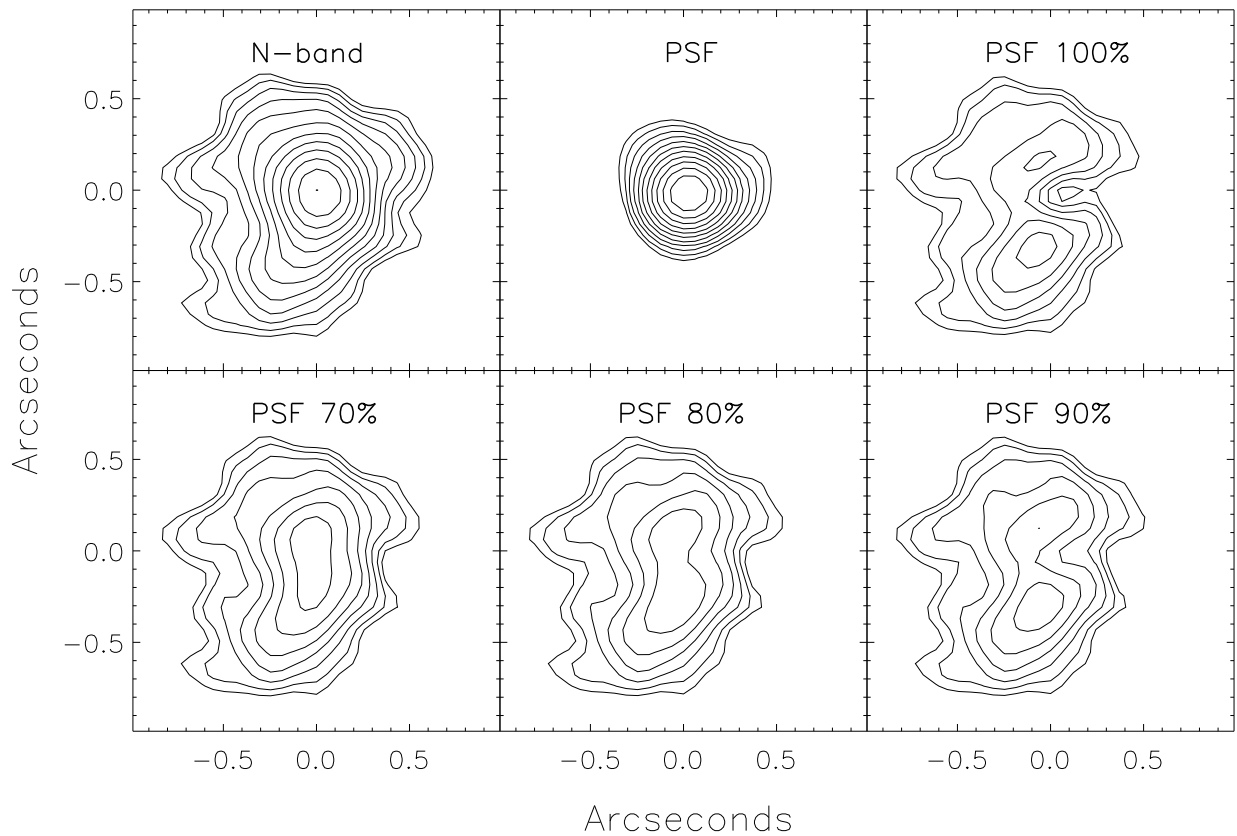}
\caption{N-band images of the central 2$\arcsec$ of Cygnus A showed with
North up and East left. All images are smoothed with a 0.18$\arcsec$
gaussian filter to enhance low level emission and scaled logarithmically.
The lowest contour represent the 4 $\protect\sigma $ level of the smoothed
data (0.067 mJy). The next image shows the PSF star $\protect\nu $ Cyg
scaled to the same level as Cygnus A for comparison. The next four images
shows the residuals of Cygnus A after subtraction of the PSF (unresolved
component) scaled to 100, 70, 80, and 90\% of the peak height. With the peak
scaled to the same height as Cygnus A (100\%), the unresolved component
represents 40\% of the total emission detected at 10 $\micron
$.}
\label{fig1}
\end{figure}

\clearpage 

\begin{figure}[tbp]
\plotone{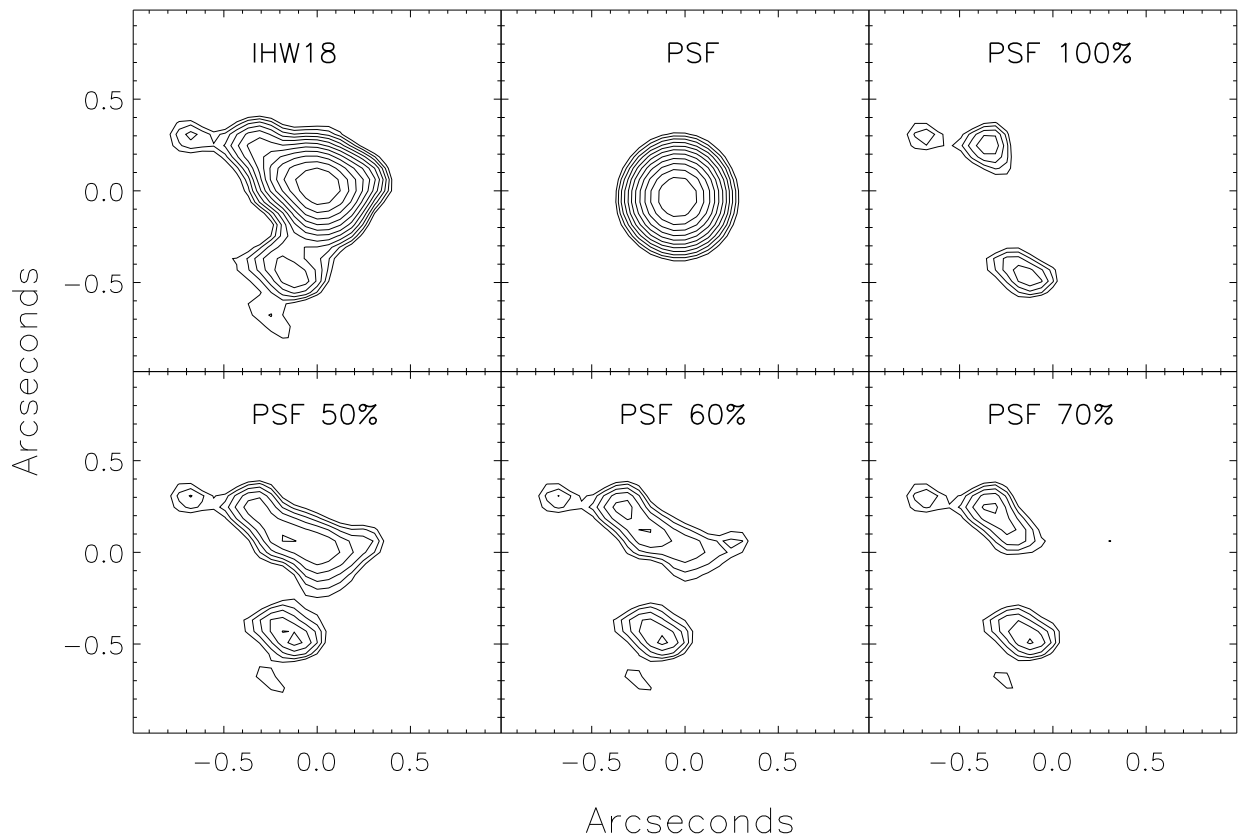}
\caption{IHW18 (18 $\micron$) images of the central 2$\arcsec$ of Cygnus A
showed with North up and East left. All images are smoothed with a 0.18$
\arcsec$ gaussian filter to enhance low level emission and are scaled
logarithmically. The lowest contour represents the 3 $\protect\sigma $ level
of the smoothed data (0.51 mJy). The next image shows the PSF star $\protect%
\gamma $ Aql scaled to the same level as Cygnus A for comparison. The next
four images show the residuals of Cygnus A after subtraction of the PSF
(unresolved component) scaled to 100, 50, 60, and 70\% of the peak height.
With the peak scaled to the same height as Cygnus A (100\%), the unresolved
component represents 60\% of the total emission detected at 18 $\micron$.}
\label{fig2}
\end{figure}

\clearpage 

\begin{figure}[tbp]
\plotone{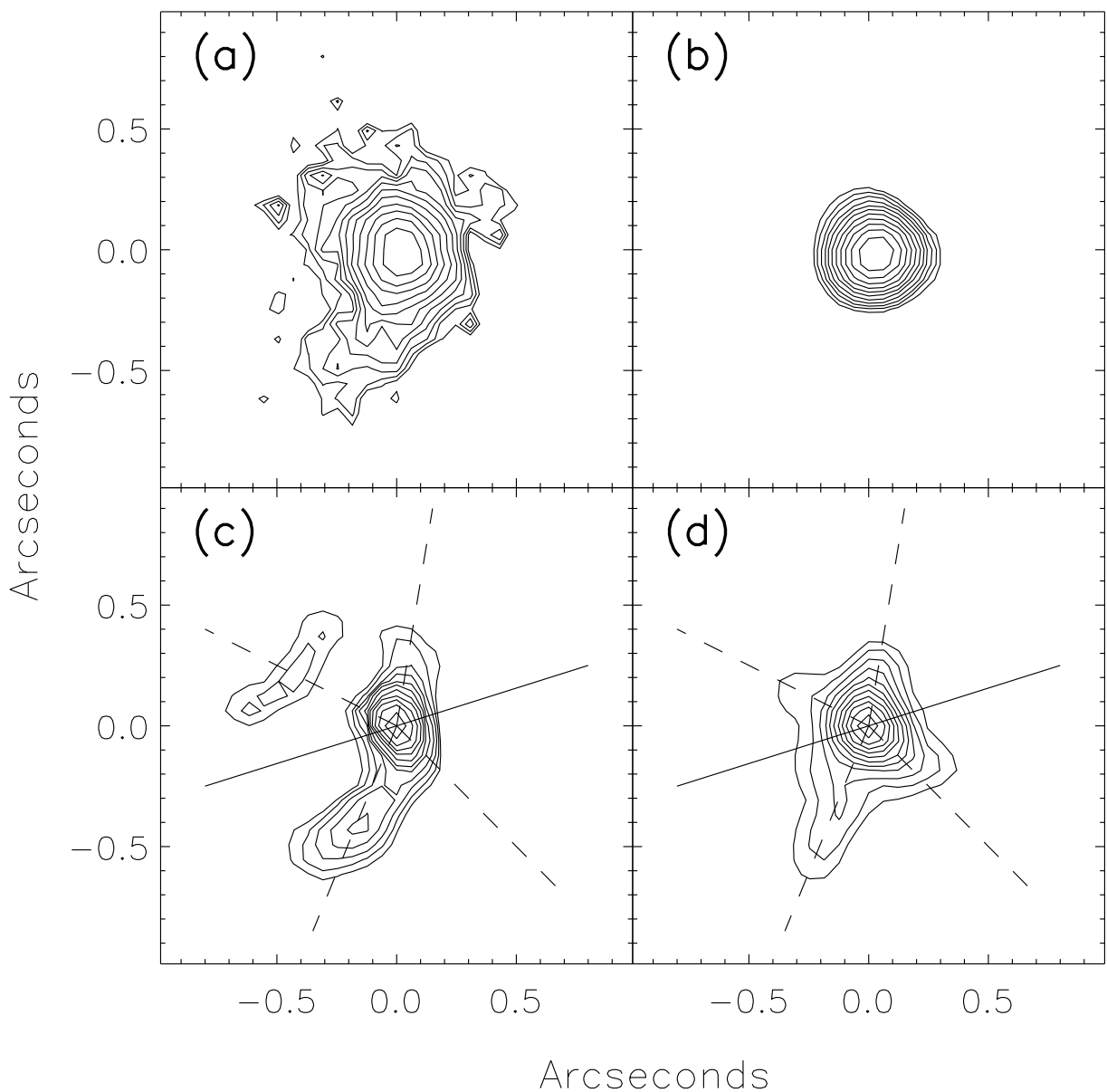}
\caption{Image (a) shows the raw unsmoothed N-Band image of Cygnus A. All
images are on a logarithmic scale. The lowest contours represent the 3 $%
\protect\sigma $ level (0.15 mJy). Image (b) is the PSF star $\protect\nu $
Cyg scaled to the same peak as image (a). Image (c) is the deconvolved image
with the lowest contours $\sim $ 2-3 $\protect\sigma $. Image (d) shows the
2.0 $\micron$ data of Tadhunter et al. (1999) scaled to show only the
nuclear morphology. Solid lines drawn in (c) and (d) show the orientation of
the radio jet axis while dashed lines are placed coincident with the
extensions seen in the 2.0 $\micron$ image and approximate the ionization
cone. }
\label{fig3}
\end{figure}

\clearpage 

\begin{figure}[tbp]
\plotone{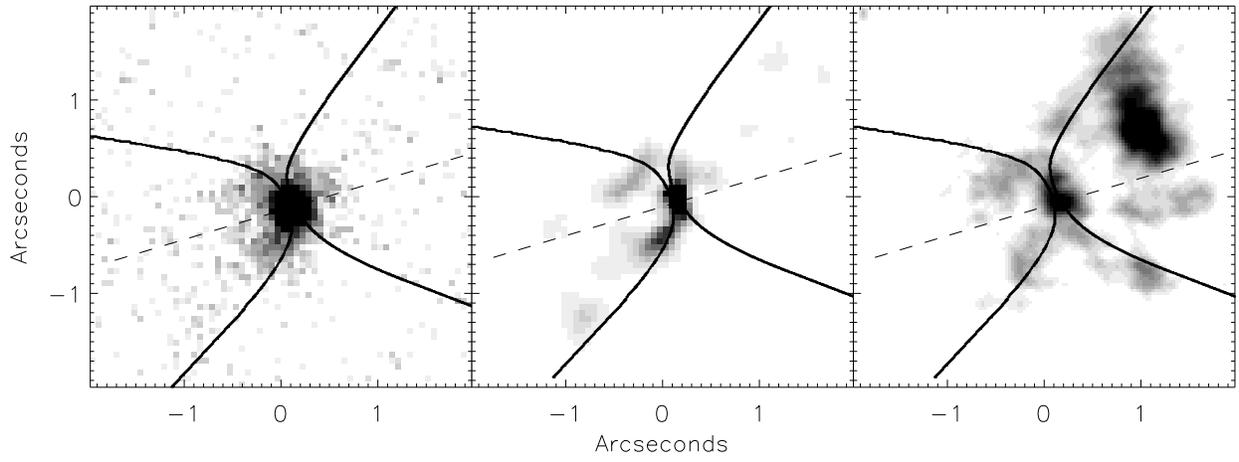}
\caption{Raw N-band data (left), followed by the deconvolved data (middle),
and the [OI] emission map from Jackson et al. (1998)(right). Solid lines
represent the parabolic ionization cone structure implied from the [OI]
emission, while the dashed line represents the radio axis of Cygnus A.}
\label{fig4}
\end{figure}

\clearpage 

\begin{figure}[tbp]
\plotone{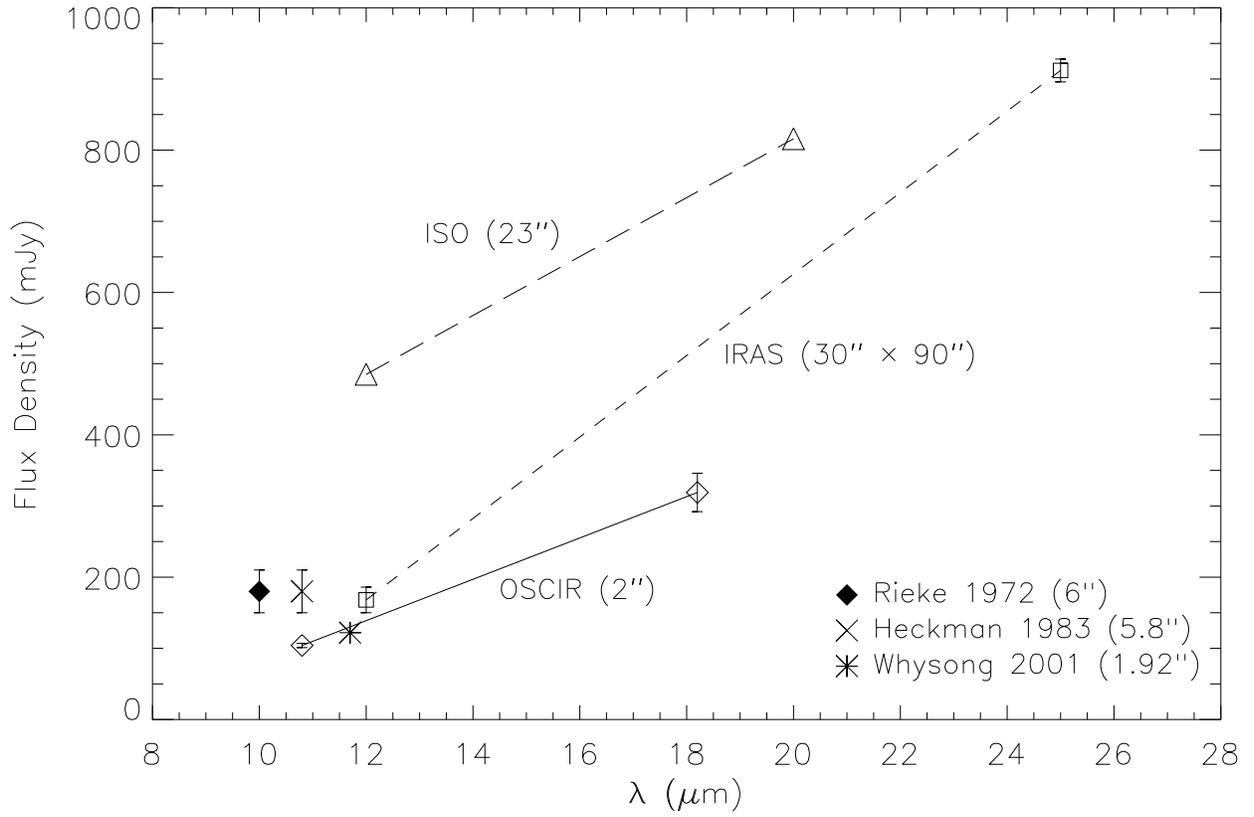}
\caption{Flux measurements as seen in Table 1 labeled with the corresponding
instrument (or reference) followed by the aperture used in parenthesis.}
\label{fig5}
\end{figure}

\clearpage 

\begin{figure}[tbp]
\plotone{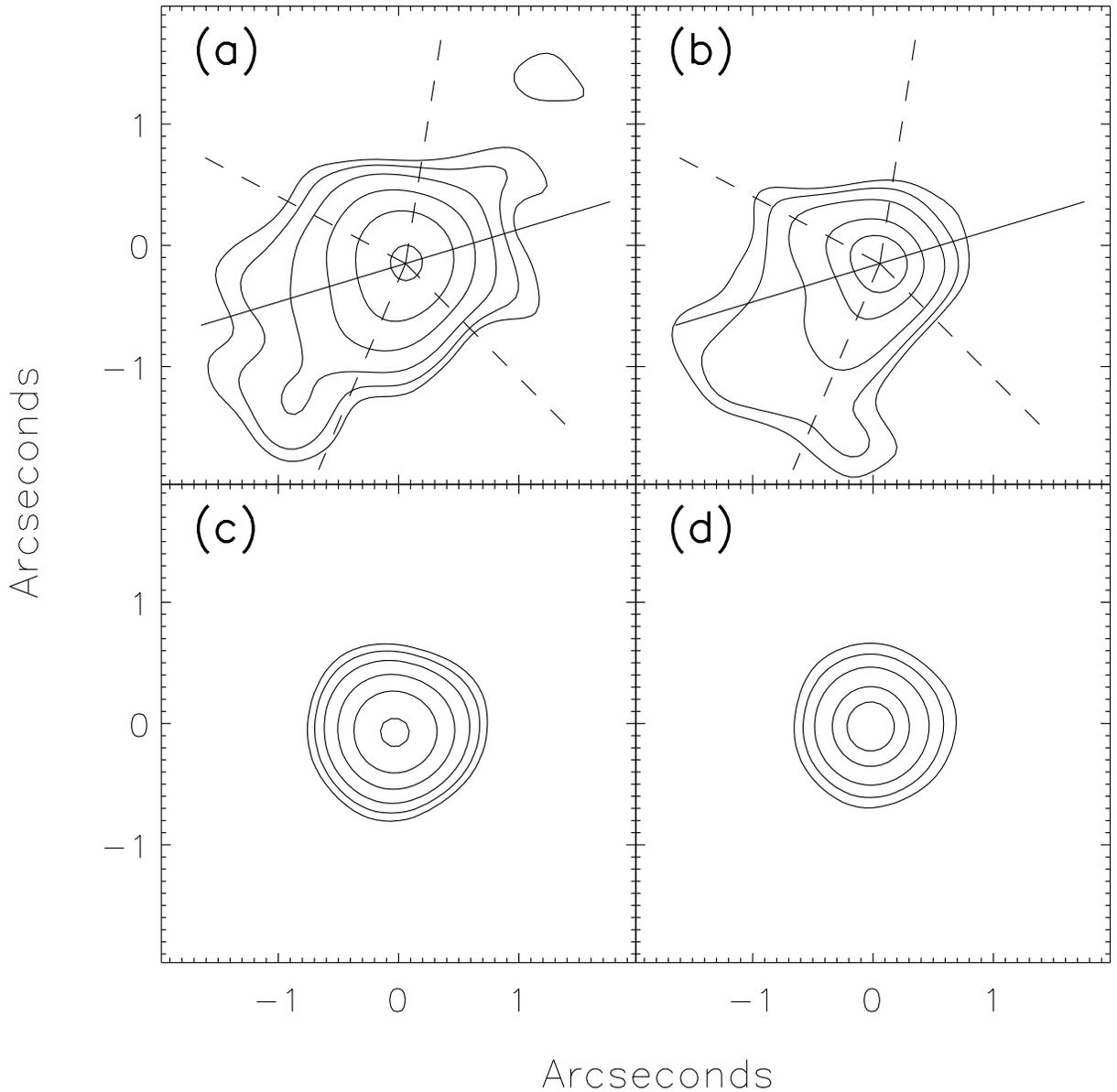}
\caption{Contours of the large-scale emission marginally detected at 10 and
18 $\micron$ respectively, smoothed by a gaussian filter of width $\sim $ 0.5%
$\arcsec$ to enhance low level emission. Image (a) shows the 10 $\micron$
emission with the lowest contours representing the 2 $\protect\sigma $
level. Subsequent contours are at 3, 5, 10, 20, and 40 $\protect\sigma $.
Image (b) shows the 18 $\micron$ emission with the lowest contours also
representing the 2 $\protect\sigma $ level. Subsequent contours are at 3, 5,
10, and 15 $\protect\sigma $. The radio axis (solid line) and linear
ionization cone structure (dashed lines) is overlaid for reference. Image
(c) shows the 10 $\micron$ PSF star $\protect\nu $ Cyg scaled the same as
(a) for comparison. Likewise image (d) shows the 18 $\micron$ PSF star $%
\protect\gamma $ Aql scaled the same as (b) for comparison.}
\label{fig6}
\end{figure}

\clearpage 

\begin{figure}[tbp]
\plotone{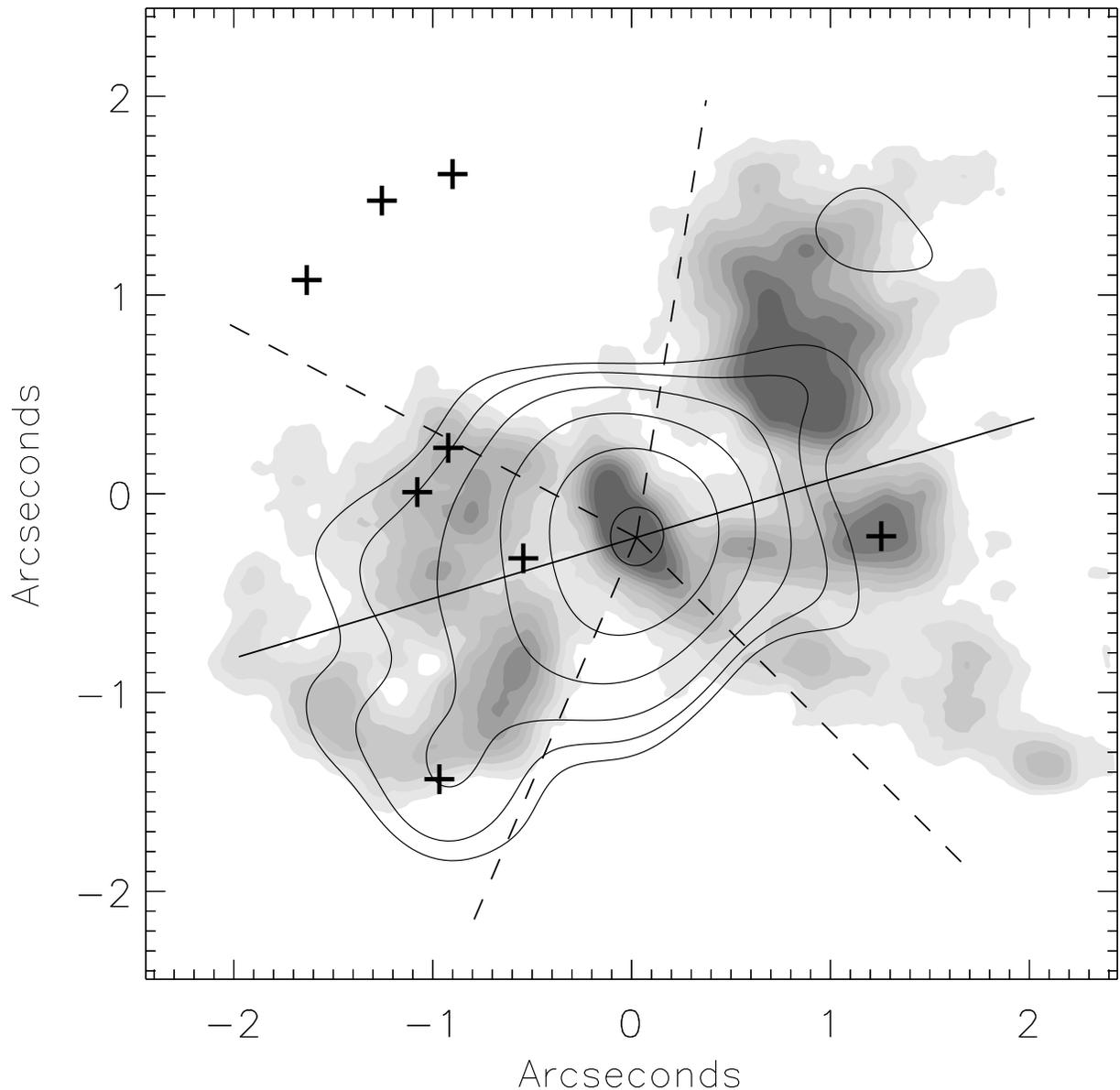}
\caption{Contours of the large-scale emission marginally detected at 10 $
\micron$ overlaid on the grayscale [OIII] map of Jackson et al. (1998). The
10 $\micron$ image is heavily smoothed by a gaussian of width $\sim $ 0.5$
\arcsec$ to enhance low level emission and centered on the radio core. The
lowest contour represents the 2 $\protect\sigma $ level, with subsequent
contours at 3, 5, 10, 20, and 40 $\protect\sigma $. Crosses indicate the
position of blue compact condensations from Jackson et al. (1998) which
possibly represent areas of star formation. }
\label{fig7}
\end{figure}

\clearpage 

\begin{figure}[tbp]
\plotone{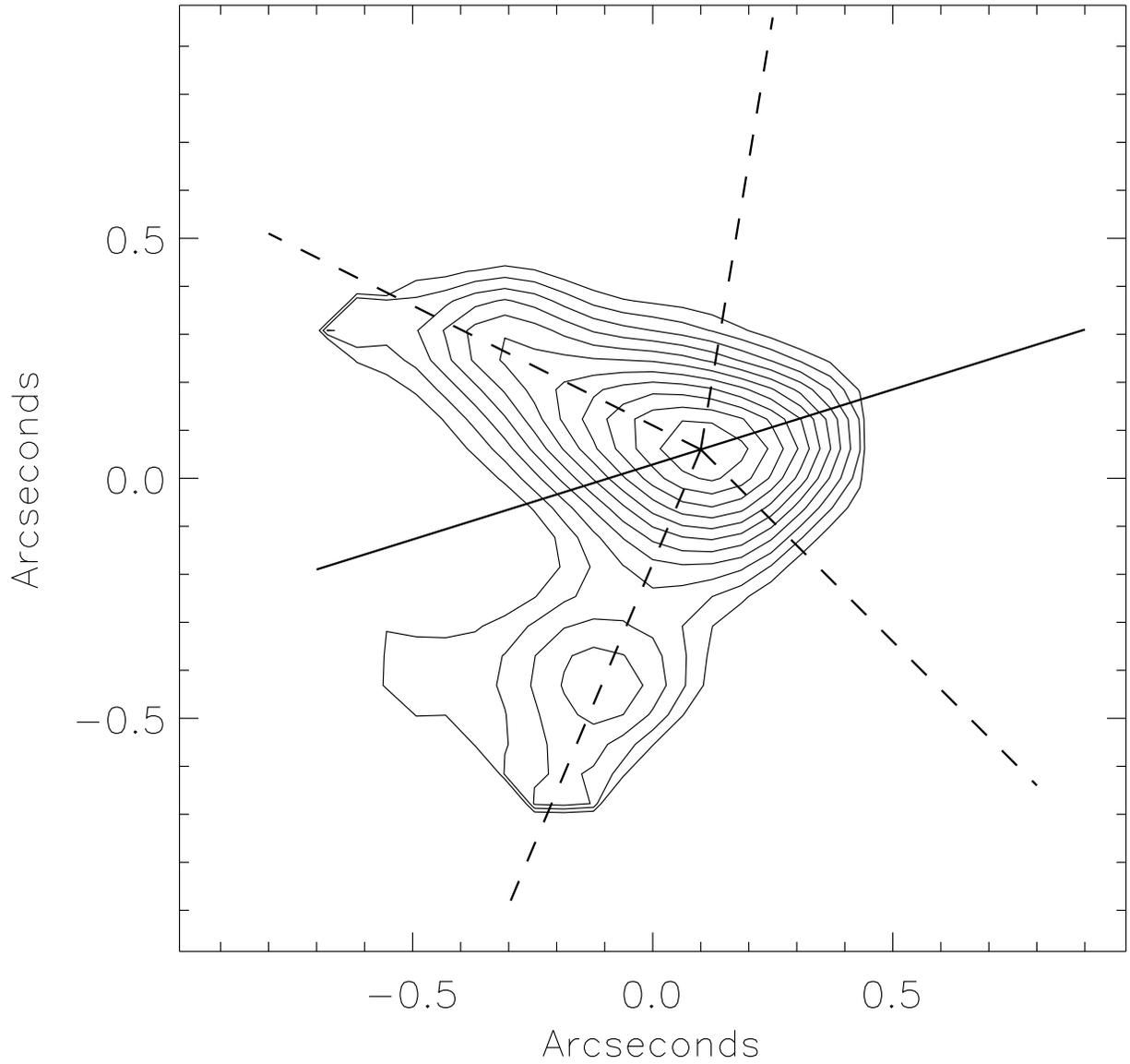}
\caption{Emission optical depth map created from the ratio of 10/18 $\micron$
emission. Contours are scaled linearly with the lowest contour at 1.0 $%
\times $ 10$^{-4}$ and the maximum at 5.9 $\times $ 10$^{-4}$. The radio
axis (solid line) and linear ionization cone structure (dashed lines)
derived from the 2.0 $\micron$ emission is overlaid for reference. }
\label{fig8}
\end{figure}


\begin{thebibliography}{Heckman, Lebofsky, Rieke, \& van Breugel(1983)}
\bibitem[Antonucci, Hurt, \& Kinney(1994)]{1994Nat...371..313A}  Antonucci,
R. R. J., Hurt, T., \& Kinney, A. 1994, \nat, 371, 313

\bibitem[Antonucci \& Miller(1985)]{1985ApJ...297..621A}  Antonucci, R.\ R.\
J.\ \& Miller, J.\ S.\ 1985, \apj, 297, 621

\bibitem[Baade \& Minkowski(1954)]{1954ApJ...119..206B}  Baade, W.\ \&
Minkowski, R.\ 1954, \apj, 119, 206

\bibitem[Cohen et al.(1995)]{1995AJ....110..275C}  Cohen, M., Witteborn, F.\
C., Walker, R.\ G., Bregman, J.\ D.\ \& Wooden, D.\ H.\ 1995, \aj, 110, 275

\bibitem[Dopita, Heisler, Lumsden, \& Bailey(1998)]{1998ApJ...498..570D}  %
Dopita, M.\ A., Heisler, C., Lumsden, S., \& Bailey, J.\ 1998, \apj, 498, 570

\bibitem[Draine \& Lee(1984)]{Draine84}  Draine, B.\ T.\ \& Lee, H.\ M.\
1984, \apj, 285, 89

\bibitem[Falcke et al.(1998)]{1998...ApJ...502...199}  Falcke, H.\ Wilson
A.\ S.\ Simpson, C.\ 1998, \apj, 502, 199

\bibitem[Haas et al.(1998)]{1998ApJ...503L.109H}  Haas, M., Chini, R.,
Meisenheimer, K., Stickel, M., Lemke, D., Klaas, U.\ \& Kreysa, E.\ 1998, %
\apjl, 503, L109

\bibitem[Heckman, Lebofsky, Rieke, \& van Breugel(1983)]%
{1983ApJ...272..400H}  Heckman, T.\ M., Lebofsky, M.\ J., Rieke, G.\ H., \&
van Breugel, W.\ 1983, \apj, 272, 400

\bibitem[Holtzman et al.(1992)]{1992AJ....103..691H}  Holtzman, J.\ A.\ et
al.\ 1992, \aj, 103, 691

\bibitem[Imanishi \& Ueno(2000)]{2000ApJ...535..626I}  Imanishi, M.\ \&
Ueno, S.\ 2000, \apj, 535, 626

\bibitem[Jackson, Tadhunter \& Sparks(1998)]{1998MNRAS.301..131J}  Jackson,
N., Tadhunter, C.\ \& Sparks, W.\ B.\ 1998, \mnras, 301, 131

\bibitem[Kawara, Nishida \& Gregory(1990)]{1990ApJ...352..433K}  Kawara, K.,
Nishida, M.\ \& Gregory, B.\ 1990, \apj, 352, 433

\bibitem[Knapp, Bies, \& van Gorkom(1990)]{1990AJ.....99..476K}  Knapp, G.\
R., Bies, W.\ E., \& van Gorkom, J.\ H.\ 1990, \aj, 99, 476

\bibitem[Lynds, O'Neil, Scowen, \& Idt(1994)]{Lynds94}  Lynds, R., O'Neil,
E.\ J., Scowen, P.\ A., \& Idt, M.\ O.\ W.\ C.\ 1994, American Astronomical
Society Meeting, 184, 4905

\bibitem[Mouri, Taniguchi, Kawara \& Nishida(1989)]{1989ApJ...346L..73M}  %
Mouri, H., Taniguchi, Y., Kawara, K.\ \& Nishida, M.\ 1989, \apjl, 346, L73

\bibitem[Ogle et al.(1997)]{1997ApJ...482L..37O}  Ogle, P.\ M., Cohen, M.\
H., Miller, J.\ S., Tran, H.\ D., Fosbury, R.\ A.\ E.\ \& Goodrich, R.\ W.\
1997, \apjl, 482, L37

\bibitem[Osterbrock \& Fulbright(1996)]{1996PASP..108..183O}  Osterbrock,
D.\ E.\ \& Fulbright, J.\ P.\ 1996, \pasp, 108, 183

\bibitem[Radomski et al.(2001)]{2001 ASP Conf. Ser. submitted}  Radomski, J.
T., Pi\~{n}a, R., Packham, C., Telesco, C., \& Tadhunter, C. 2001, ASP
Conf. Ser. 249: The Central Kiloparsec of Starbursts and AGN: the La Palma
Connection, ed. J.H. Knapen, J.E. Beckman, I. Shlosman \& T.J. Mahoney, 325

\bibitem[Rieke \& Low(1972)]{1972ApJ...176L..95R}  Rieke, G.\ H.\ \& Low,
F.\ J.\ 1972, \apjl, 176, L95

\bibitem[Schmitt \& Kinney(1996)]{1996ApJ..463..498}  Schmitt, H.\ R.\ \&
Kinney, A.\ L.\ 1996, \apj, 463, 498

\bibitem[Smith, Aitken \& Roche(1989)]{1989MNRAS.241..425S}  Smith, C.\ H.,
Aitken, D.\ K.\ \& Roche, P.\ F.\ 1989, \mnras, 241, 425

\bibitem[Stockton, Ridgway \& Lilly(1994)]{1994AJ....108..414S}  Stockton,
A., Ridgway, S.\ E.\ \& Lilly, S.\ J.\ 1994, \aj, 108, 414

\bibitem[Tadhunter et al.(1999)]{1999ApJ...512L..91T}  Tadhunter, C.\ N.,
Packham, C., Axon, D.\ J., Jackson, N.\ J., Hough, J.\ H., Robinson, A.,
Young, S.\ \& Sparks, W.\ 1999, \apjl, 512, L91

\bibitem[Tadhunter et al.(2000)]{2000MNRAS.313L..52T}  Tadhunter, C.\ N.\ et
al.\ 2000, \mnras, 313, L52

\bibitem[Telesco(1988)]{1988ARA&A..26..343T}  Telesco, C.\ M.\ 1988, \araa,
26, 343

\bibitem[Terlevich, Diaz, \& Terlevich(1990)]{1990MNRAS.242..271T}  %
Terlevich, E., Diaz, A.\ I., \& Terlevich, R.\ 1990, \mnras, 242, 271

\bibitem[Thornton, Stockton \& Ridgway(1999)]{1999AJ....118.1461T}  %
Thornton, R.\ J., Stockton, A.\ \& Ridgway, S.\ E.\ 1999, \aj, 118, 1461

\bibitem[Tresch-Fienberg et al.(1987)]{1987ApJ...312..542T}  %
Tresch-Fienberg, R., Fazio, G.\ G., Gezari, D.\ Y., Lamb, G.\ M., Shu, P.\
K., Hoffmann, W.\ F., \& McCreight, C.\ R.\ 1987, \apj, 312, 542

\bibitem[Ueno et al.(1994)]{1994ApJ...431L...1U}  Ueno, S., Koyama, K.,
Nishida, M., Yamauchi, S.\ \& Ward, M.\ J.\ 1994, \apjl, 431, L1

\bibitem[Velusamy \& Langer(1998)]{1998Natur.392..685V}  Velusamy, T.\ \&
Langer, W.\ D.\ 1998, \nat, 392, 685

\bibitem[Ward, Blanco, Wilson \& Nishida(1991)]{1991ApJ...382..115W}  Ward,
M.\ J., Blanco, P.\ R., Wilson, A.\ S.\ \& Nishida, M.\ 1991, \apj, 382, 115

\bibitem[Whysong \& Antonuuci (2001))]{2001 astro-ph/0106381}  Whysong, D.
\& Antonucci, R. 2001, ApJL, submitted (astro-ph/0106381)

\bibitem[Young et al.(1999)]{1999MNRAS.303..227Y}  Young, S., Corbett, E.\
A., Giannuzzo, M.\ E., Hough, J.\ H., Robinson, A., Bailey, J.\ A.\ \& Axon,
D.\ J.\ 1999, \mnras, 303, 227
\end{thebibliography}
\end{document}